\documentclass{JAC2001}
\usepackage{graphicx}
\newcommand{\ud}{\mathrm{d}}
\begin{document}

\title{MODELLING OF BEAM-BEAM EFFECTS IN MULTISCALES}
\author{Antonina  N. Fedorova,  Michael  G. Zeitlin \\
IPME, RAS, V.O. Bolshoj pr., 61, 199178, St.~Petersburg, Russia
\thanks{e-mail: zeitlin@math.ipme.ru}\thanks{ http://www.ipme.ru/zeitlin.html;
http://www.ipme.nw.ru/zeitlin.html}   
}
\maketitle

\begin{abstract}
We present the applications of nonlinear local harmonic analysis methods       
to the modelling of beam-beam                                  
interaction. Our approach                                                      
is based on methods provided the possibility to work with dynamical                
beam localization in phase space.                                              
The consideration of Fokker-Planck or Vlasov-Maxwell models is based on                
a number of anzatzes, which reduce initial problems to a number of             
dynamical systems (with constraints) and on variational-wavelet approach       
to polynomial/rational approximations for reduced nonlinear dynamics.          
We calculate contribution to full dynamics (partition function)       
from all underlying subscales via nonlinear eigenmodes decomposition.                                                               
\end{abstract}

\section{Introduction}
In this paper we consider the applications of nu\-me\-ri\-cal\--analytical 
technique based on the methods of local nonlinear harmonic
analysis or wavelet analysis to nonlinear models of 
beam-beam interactions which can be characterized by collective type behaviour.
We consider different but related models [1] of the beam-beam interaction from the point of view of
construction of reasonable numerical-analytical approaches. It is very important because some beam-beam effects
limit luminosity and stability of collider. 
Such approaches may be useful in all models in which  it is 
possible and reasonable to reduce all complicated problems related with 
statistical distributions to the problems described 
by systems of nonlinear ordinary/partial differential 
equations with or without some (functional) constraints.
Wavelet analysis is a set of mathematical
methods, which gives us the possibility to work with well-localized bases in
functional spaces and gives the maximum sparse forms for the general 
type of operators (differential, integral, pseudodifferential) in such bases. 
Our approach is based on the 
variational-wavelet approach from [2]-[13],
which allows us to consider polynomial and rational type of 
nonlinearities.
The constructed solution has the multiscale/multiresolution decomposition via 
nonlinear high-localized eigenmodes. 
In this way we give contribution to our full solution
from each scale of resolution or each time/(phase) space scale or from each nonlinear eigenmode. 
Fig.1 demonstrates such decomposition for the finite kick term.
The same is correct for the contribution to power spectral density
(energy spectrum): we can take into account contribution from each
level/scale of resolution.
In all models below numerical modelling demonstrates the appearance of (coherent) high-localized structures
and (stable) pattern formation.
Starting  in part 2 from beam-beam interaction models [1]
we consider in part 3 the approach based on
variational-wavelet formulation. 
We give explicit representation for all dynamical variables in the base of
compactly supported wavelets or nonlinear eigenmodes.  Our solutions
are parametrized
by solutions of a number of reduced algebraical problems one from which
is nonlinear with the same degree of nonlinearity as initial models and the rest  are
the linear problems which correspond to concrete
details of wavelet calculations. 
In part 4 we consider numerical modelling based on our analytical approach.

\section{BEAM-BEAM MODELLING}

In A.~Chao e.a. model [1] for the simulation of beam-beam interaction the initial 
collective description by some sort of Vlasov-Maxwell equation for distribution function $f(s,x,p)$
\begin{eqnarray}
\frac{\partial f}{\partial s}+p\frac{\partial f}{\partial x}-
  \Big(k(s)x-F(x,s,f)\Big)\frac{\partial f}{\partial p}=0
\end{eqnarray}
is reduced to Fokker-Planck (FP) equation on the first stage

\begin{eqnarray}
&&\frac{\partial f_k}{\partial s}-\Gamma _kf_k+p\frac{\partial f_k}{\partial x}-\\
&&\Big(F_k(s,x,f)-G_k(s,p)\Big)\frac{\partial f_k}{\partial p}
=D\frac{\partial^2 f_k}{\partial p^2}\nonumber
\end{eqnarray}
and to
some nontrivial dynamical system with complex behaviour
\begin{eqnarray}
&&\frac{\ud^2\sigma_k}{\ud s^2}+ \Gamma_k\frac{\ud\sigma_k}{\ud s}+
 F_k\sigma_k=\frac{1}{\beta^2_ka^2_k\sigma^3_k}\nonumber\\
&&\frac{\ud a_n}{\ud s}=\Gamma_ka_k(1-a_k^2\sigma^2_k)
\end{eqnarray}
on the second stage.
Its solution gives the parameters of enveloping 
gaussian anzatz for solution of FP equation.
Related model of R.~Davidson e.a.[1] is based on Vlasov-Maxwell equations:

\begin{eqnarray}
&&\frac{\partial f_k}{\partial s}+\nu_k p\frac{\partial f_k}{\partial x}-
  \frac{\partial H_k}{\partial x}\frac{\partial f_k}{\partial p}=0\\
&&H_k=\frac{\nu_k}{2}(p^2+x^2)+\lambda_k\delta_p(s) V_k(x,s)\nonumber\\
&&\frac{\partial^2V_k}{\partial x^2}=4\pi\int\ud pf_{3-k}(x,p,s)\nonumber
\end{eqnarray}

\begin{figure}[htb]
\centering
\includegraphics*[width=65mm]{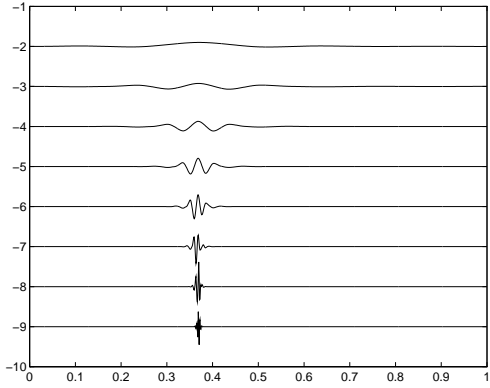}
\caption{Finite kick decomposition.}
\end{figure}

\section{VARIATIONAL MULTISCALE REPRESENTATION}

We obtain our multiscale\-/mul\-ti\-re\-so\-lu\-ti\-on representations (formulae (9) below) for solutions of equations
(1),(2),(4) via variational-wavelet approach for the following formal systems of equations (with the
corresponding obvious constraints on distribution function), which is the general form of these equations. 

Let L be an arbitrary (non) li\-ne\-ar dif\-fe\-ren\-ti\-al\-/\-in\-teg\-ral operator with matrix dimension $d$, 
which acts on some set of functions
$\qquad\Psi\equiv\Psi(s,x,p)=\Big(\Psi^1(s,x,p),\dots,$\\*$\Psi^d(s,x,p)\Big)$, $ s,x,p\in\Omega\subset{\bf R}^3$
from $L^2(\Omega)$:
\begin{equation}
L\Psi\equiv L(Q,s,x,p)\Psi(s,x,p)=0,
\end{equation}
where
\begin{eqnarray}
&&Q\equiv Q_{d_1,d_2,d_3}(s,x,p,\partial /\partial s,\partial /\partial x,\partial /\partial p)=\\
&&\sum_{i_1,i_2,i_3=1}^{d_1,d_2,d_3}
a_{i_1i_2i_3}(s,x,p)
\Big(\frac{\partial}{\partial s}\Big)^{i_1}\Big(\frac{\partial}{\partial x}\Big)^{i_2}\Big(\frac{\partial}{\partial p}\Big)^{i_3}\nonumber
\end{eqnarray}
Let us consider now the N mode approximation for solution as the following ansatz (in the same way
we may consider different ansatzes):
\begin{equation}
\Psi^N(s,x,p)=\sum^N_{r,s,k=1}a_{rsk}A_r(s)\otimes B_s(x)\otimes C_k(p)
\end{equation}
We shall determine the coefficients of expansion from the following conditions
(different related variational approaches are considered in [2]-[13]):
\begin{equation}
\ell^N_{k\ell m}\equiv\int(L\Psi^N)A_k(s)B_\ell(x)C_m(p)\ud s\ud x\ud p=0
\end{equation}
So, we have exactly $dN^3$ algebraical equations for  $dN^3$ unknowns $a_{rsk}$.
Such variational approach reduces the initial problem to the problem of solution 
of functional equations at the first stage and some algebraical problems at the second
stage. We consider the multiresolution expansion as the second main part of our 
construction. 
The solution is parametrized by solutions of two set of reduced algebraical
problems, one is linear or nonlinear
(depends on the structure of operator L) and the rest are some linear
problems related to computation of coefficients of algebraic equations (8).
These coefficients can be found  by some wavelet methods
by using
compactly supported wavelet basis functions for expansions (7).
We may consider also different types of wavelets including general wavelet packets [2]-[13].
The constructed solution has the following mul\-ti\-sca\-le\-/\-mul\-ti\-re\-so\-lu\-ti\-on decomposition via 
nonlinear high\--\-lo\-ca\-li\-zed eigenmodes 
{\setlength\arraycolsep{0pt}
\begin{eqnarray}\label{eq:z}
&&f(s,x,p)=\sum_{(i,j,k)\in Z^3}a_{ijk}A^i(s)B^j(x)C^k(p),\\
&&A^i(s)=A_N^{i,slow}(s)+\sum_{r\geq N}A^i_r(\omega^1_rs),\ \omega^1_r\sim 2^r \nonumber\\
&&B^j(x)=B_M^{j,slow}(x)+\sum_{l\geq M}B^j_l(\omega^2_lx),\ \omega^2_l\sim 2^l \nonumber\\
&&C^k(p)=C_K^{k,slow}(p)+\sum_{m\geq K}C^k_m(\omega^3_mp),\ \omega^3_m\sim 2^m \nonumber
\end{eqnarray}}
which corresponds to the full multiresolution expansion in all underlying time/space 
scales.
Formula (\ref{eq:z}) gives us expansion into the slow part $f_{N,M,K}^{slow}$
and fast oscillating parts for arbitrary N, M, K.  So, we may move
from coarse scales of resolution to the 
finest one to obtain more detailed information about our dynamical process.
The first terms in the RHS of formulae (9) correspond on the global level
of function space decomposition to  resolution space and the second ones
to detail space.
Particular one-dimensional case of formulae (9) determines the solution of
equations (3) (more exactly corresponding constructions are considered in other papers presented during this Conference).
But, it should be noted that in this one-dimensional case we have really nontrivial nonlinear dynamics only for functional parameters of enveloping gaussians,
which give the solution only for linearization of equations (1),(2),(4).
As we demonstrated, our representation (9) provides the solution as in linear as in nonlinear cases without any perturbation technique
but on the level of expansions in (functional) space of solutions.
The using of wavelet basis with high-localized properties provides good convergence properties of constructed solution (9). 
Because affine
group of translation and dilations is inside the approach, this
method resembles the action of a microscope. We have contribution to
final result from each scale of resolution from the whole
infinite scale of spaces or from each underlying scale: the closed subspace
$V_j (j\in {\bf Z})$ corresponds to  level j of resolution, or to scale j.
Our full multiresolution/multiscale decomposition of functional space
$L^2 ({\bf R}^n)$ of solutions of initial problems,
which is a sequence of increasing closed subspaces $V_j$:
\noindent
$\dots V_{-2}\subset V_{-1}\subset V_0\subset V_{1}\subset V_{2}\subset\dots$,
provides us with formulae (9).
This functional space decomposition corresponds to (exact) nonlinear
eigenmode decomposition.
It should be noted that such representations 
give the best possible localization
properties in the corresponding (phase)space/time coordinates. 
In contrast with different approaches, formulae (9) does not use perturbation
technique or linearization procedures 
and represents dynamics via generalized nonlinear localized eigenmodes expansion.  
So, by using wavelet bases with their good (phase)space/time      
localization properties we can construct high-localized coherent structures in      
spa\-ti\-al\-ly\--ex\-te\-nd\-ed stochastic systems with collective behaviour.

\begin{figure}[htb]
\centering
\includegraphics*[width=65mm]{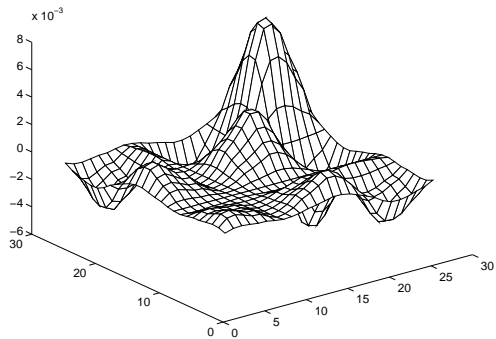}
\caption{Eigenmode of level 1.}
\end{figure}

\begin{figure}[htb]
\centering
\includegraphics*[width=65mm]{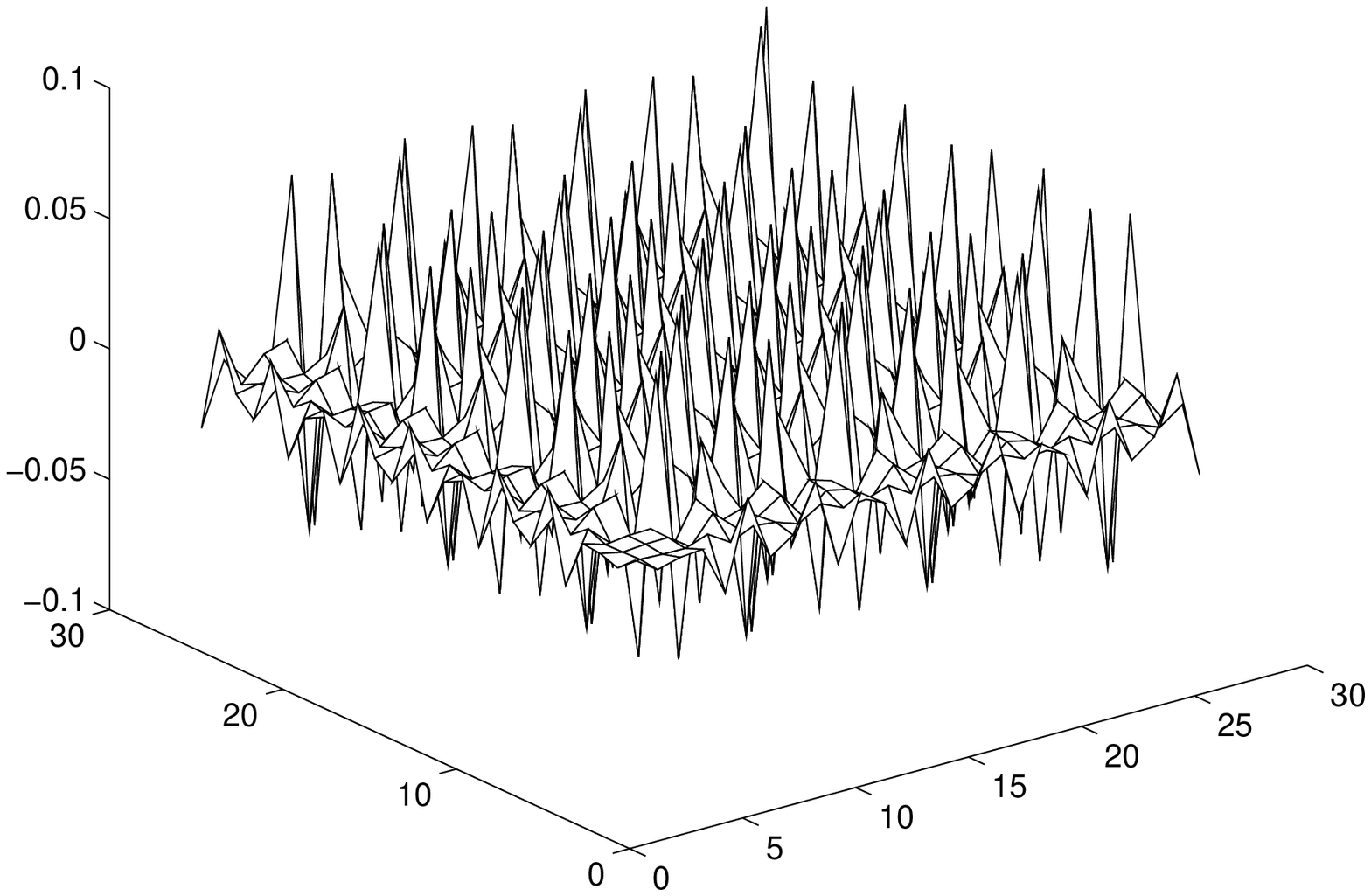}
\caption{Eigenmode of level 4.}
\end{figure}

\begin{figure}[htb]
\centering
\includegraphics*[width=65mm]{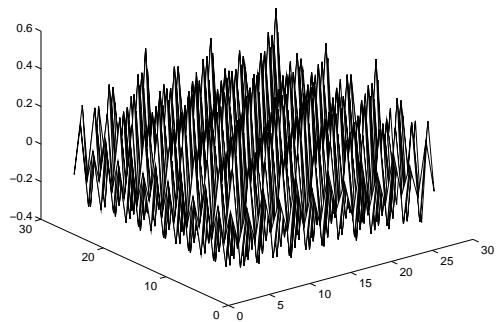}
\caption{Six-eigenmodes representation.}
\end{figure}

\newpage
\section{MODELLING}

Multiresolution/multiscale representations for solutions of equations from part 2
in the high-localized ba\-ses\-/\-ei\-gen\-mo\-des
are demonstrated on Fig.~2--Fig.~4.
On Fig.~2,  Fig.~3 we present contribution to the full expansion (9) from level 1 and level 4
of full decomposition. Fig. 4 show the representation for full solution, constructed
from the first 6 eigenmodes (6 levels in formula (9)). 

\section{ACKNOWLEDGMENTS}

We would like to thank The U.S. Civilian Research \& Development Foundation (CRDF) for
support (Grants TGP-454, 455), which gave us the possibility to present our nine papers during
PAC2001 Conference in Chicago and Ms.Camille de Walder from CRDF for her help and encouragement.


\begin{thebibliography}{13}

\bibitem{1}
A. Chao, e.a., Los Alamos preprint, physics/0010055.
R.C. Davidson, e.a., Los Alamos preprint, physics/0104086.

\bibitem{2}
A.N. Fedorova and M.G. Zeitlin, 
 {\it Math. and Comp. in Simulation}, {\bf 46}, 527, 1998.

\bibitem{3}
A.N. Fedorova and M.G. Zeitlin,
{\it New Applications of Nonlinear and Chaotic Dynamics in Mechanics}, 31, 101
Klu\-wer,  1998.

\bibitem{4}
A.N. Fedorova and M.G. Zeitlin,
{\bf CP405}, 87, American Institute of Physics, 1997.\\
Los Alamos preprint, physics/9710035.

\bibitem{5}
A.N. Fedorova, M.G. Zeitlin and Z.~Parsa, 
Proc. PAC97 
{\bf 2}, 1502, 1505, 1508, APS/IEEE, 1998.

\bibitem{6}
A.N. Fedorova, M.G. Zeitlin and Z.~Parsa, 
Proc. EPAC98, 930, 933, Institute of Physics, 1998.

\bibitem{7}
A.N. Fedorova, M.G. Zeitlin and Z.~Parsa,    
{\bf CP468}, 48, American Institute of Physics, 1999.
Los Alamos preprint, physics/990262.

\bibitem{8}
A.N. Fedorova, M.G. Zeitlin and Z.~Parsa,  
{\bf CP468}, 69, American Institute of Physics, 1999.
Los Alamos preprint, physics/990263.

\bibitem{9}
A.N. Fedorova and M.G. Zeitlin,  
Proc. PAC99, 
1614, 1617, 1620, 2900, 2903,
2906, 2909, 2912, APS/IEEE, New York, 1999.\\
Los Alamos preprints: 
physics/9904039, 9904040,\\ 9904041, 9904042, 9904043, 
9904045, 9904046, 9904047.

\bibitem{10}
A.N. Fedorova and M.G. Zeitlin,
The Physics of High Brightness Beams, 235, World Scientific, 2000. 
Los Ala\-mos pre\-pri\-nt: physics/0003095.

\bibitem{11}
A.N. Fedorova and M.G. Zeitlin,  Proc. EPAC00, 
415, 872,  1101, 1190, 1339, 2325,Austrian Acad.Sci.,2000.\\ 
Los Alamos preprints: physics/0008045, 0008046,\\
 0008047, 0008048, 0008049, 0008050.

\bibitem{12}
A.N. Fedorova, M.G. Zeitlin, Proc. 20 International Linac Conf., 300, 
303, SLAC, Stan\-ford, 2000. 
Los Ala\-mos pre\-pri\-nts: physics/0008043, 0008200.

\bibitem{13}
A.N. Fedorova, M.G. Zeitlin, Los Ala\-mos pre\-prints:\\
 physics/0101006, 0101007
and World Scientific, in press. 

\end{thebibliography}
\end{document}